# Characterization of the probability and information entropy of a process with an increasing sample space by different functional forms of expansion, with an application to hyperinflation

Laurence Francis Lacey

Lacey Solutions Ltd, Skerries, County Dublin, Ireland

# Characterization of the probability and information entropy of a process with an increasing sample space by different functional forms of expansion, with an application to hyperinflation

Laurence Francis Lacey

Lacey Solutions Ltd, Skerries, County Dublin, Ireland

## Abstract

There is a random variable (X) with a determined outcome (i.e., $X = x_0$), $p(x_0) = 1$. Consider $x_0$ to have a discrete uniform distribution over the integer interval [1, s], where the size of the sample space (s) = 1, in the initial state, such that $p(x_0) = 1$. What is the probability of $x_0$ and the associated information entropy (H), as s increases by means of different functional forms of expansion? Such a process has been characterised in the case of (1) a mono-exponential expansion of the sample space; (2) a power function expansion; (3) double exponential expansion. The double exponential expansion of the sample space with time (from a natural log relationship between t and n) describes a "hyperinflationary" process. Over the period from the middle of 1920 to the end of 1923, the purchasing power of the Weimar Republic paper Mark to purchase one gold Mark became close to zero (1 paper Mark = $10^{-12}$ gold Mark). From the purchasing power of the paper Mark to purchase one gold Mark, the information entropy of this hyperinflationary process was determined.

## 1. Introduction

Information entropy (H) will be used as the measure of entropy and in the following form:

$$H = -\sum_{i}^{n} p_i \, log_e \, p_i$$

H = 0 for a random variable (X) with a determined outcome (i.e., X = $x_0$), p($x_0$) = 1 [1].

The primary objective of this paper is to characterise the probability and information entropy of a process, in which the sample space (s) increases by means of different functional forms of expansion.

## 2. Methods

### 2.1 Expansion process

There is a random variable (X) with a determined outcome (i.e., X = $x_0$), p($x_0$) = 1. Consider $x_0$ to have a discrete uniform distribution over the integer interval [1, s], where the size of the sample space (s) = 1, in the initial state, such that p($x_0$) = 1, H=0.

For a sample space of size, $s_0 > 1$, let $x_0$ have a discrete uniform distribution over the integer interval $[1, s_0]$, giving:

$$p(x_0) = \frac{1}{s_0}$$

where, $p(x_0)$ is the probability mass function over the integer interval $[1, s_0]$. The cumulative probability of $p(x_0)$ over $[1, s_0]$ is equal to 1. Let $p(x_0|s_0)$ be the <u>cumulative probability</u> of $p(x_0)$ over $[1, s_0]$, i.e.,

$$p(x_0|s_0) = 1$$

Double the sample space in which $x_0$ will be uniformly distributed, i.e., $s_1 = 2 \text{ x } s_0$, and partition the new sample space into 2 equally sized, mutually exclusive sample spaces, each of size $s_0$. Now:

$$p(x_0) = \frac{1}{s_1} = \frac{1}{2 \text{ x } s_0}$$

The cumulative probability of $x_0$, in any <u>one</u> of the two partitions of size $s_0$, is now:

$$p(x_0|s_1) = ½$$

Repeating the process a second time, the cumulative probability of $x_0$, in any <u>one</u> of the four partitions, each of size $s_0$, is now:

$$p(x_0|s_2) = ¼$$

Repeating the process n times, the cumulative probability, for $x_0$ in any <u>given</u> partition of size $s_0$ is:

$$p(x_0|s_n) = (½)^n$$

This is an exponential process, with the sample space expanding at an exponential rate, with:

$$\mathrm{s(n)} = 2^{\mathrm{n}} = \exp(log_e 2 \text{ x n})$$

The corresponding probability and information entropy of the system, as a function of n, is given by:

$$\mathrm{p}(x_0|s_n) = \exp(-log_e 2 \text{ x n})$$

$$H(n) = -2^n x\, (½)^{\mathrm{n}}\, x\, log_e(½)^{\mathrm{n}} \;\; = n\, x\, log_e 2$$

## 2.2 Time-dependent expansion processes

Three different time-dependent expansion processes will be described:

(1) **Exponential time-dependent expansion:** If the expansion occurs at an exponential rate (rate constant = λ) with time (t), such that:

$$t \;=\; n\, x \frac{log_e(2)}{\lambda}, \text{giving } n = \frac{\lambda \text{ x t}}{log_e(2)}$$

By substituting t for n, gives:

$$\mathrm{s(t)} = \exp(\lambda \text{ x t})$$

$$\mathrm{p}(x_0|t) = \exp(-\lambda \text{ x t})$$

$$H(\mathrm{t}) = \lambda \text{ x t}$$

(2) **Power time-dependent expansion:** If the expansion occurs at a rate with time (t) such that:

$$t = 2^n, \text{giving } n = \frac{log_e(t)}{log_e(2)}$$

By substituting t for n, gives

$$\mathrm{s(t)} = \mathrm{t}$$

$$\mathrm{p}(x_0|t) = \mathrm{t}^{-1}$$

$$H(\mathrm{t}) = log_e(t)$$

This is the characterization of a power function expansion and this functional form of expansion can be show for any expansion in which:

$$\mathrm{t}^a = 2^n, \text{giving } \mathrm{t} = 2^{n^{1/a}} \ and \ n = \frac{a\ x\ log_e(t)}{log_e(2)} \text{ for a > 0.}$$

In this case, substituting t for n, gives:

$$\mathrm{s(t)} = \mathrm{t^a}$$

$$\mathrm{p}(x_0|t) = \mathrm{t^{-a}}$$

$$H(\mathrm{t}) = log_e(t^{\mathrm{a}})$$

(3) **Double exponential time-dependent expansion:** If the expansion occurs at a rate with time (t) such that:

$$t \ = \lambda^{-1}\ x\ log_e\left(\frac{n}{a}\right), \text{giving}\ n = a\ x\ \exp\ (\lambda\ x\ t)$$

for a > 0.

In this case, substituting t for n, gives:

$$\mathrm{s(t)} = \ \exp\ (log_e(2^{\mathrm{a}})\ x\ \exp\ (\lambda\ x\ t))$$

$$\mathrm{p}(x_0|t) = \exp\ (log_e(2^{\mathrm{a}})\ x\ \exp\ (-\lambda\ x\ t))$$

$$\mathrm{H(t)} = log_e(2^{\mathrm{a}})\ x\ \exp\ (\lambda\ x\ t)$$

## 2.3 Time-dependent expansion via three (or more) simultaneous, independent processes

If the expansion of the sample space occurs simultaneously via three (or more) independent processes (P1, P2, P3) with time, the combined overall process can be determined by the following equations:

$$s(t) = s\big(P1(t)\big)\ x\ s(P2(t))\ x\ s(P3(t))$$

$$\mathrm{p}(x_0|t) = \ \mathrm{p}(x_0|P1, t)\ x\ \mathrm{p}(x_0|P2, t)\ x\ \mathrm{p}(x_0|P3, t)$$

$$H(P(t)) = H(P1(t)) + H(P2(t)) + H(P3(t))$$

All results of the data analysis given below were obtained using Microsoft Excel 2019, 32-bit version.

# 3. Results

## 3.1 Different functional forms of expansion

Three different functional forms of expansion of the sample space have been characterised and be summarised in Table 1.

Table 1: Three different functional forms of expansion of the sample space

| **Type of Expansion of the sample space with time** | **Relationship between t (time) and n (no. of sample space doublings)** | **Functional form of $\mathbf{p}(x_0\|t)$** | **Functional form of $\mathbf{H}(\mathbf{t})$** |
|---|---|---|---|
| **Exponential** | Linear | Exponential | Linear |
| **Power** | Exponential | Power | Natural log |
| **Double exponential** | Natural log | Double exponential | Exponential |

Examples are given in which the three types of sample space expansion are compared in terms of (1) $\mathrm{p}(x_0|t)$ (Figure 1); and (2) $\mathrm{H(t)}$ (Figure 2), respectively. Please note that time rescaling was required in order to compare the probabilities and associated entropies of the different processes at the same time-point, t. For an exponential expansion process, $\mathrm{p}(x_0|t) = 1$, when t=0; for a power expansion, $\mathrm{p}(x_0|t) = 1$, when t=1; and for the double exponential expansion (from a natural log relationship between t and n), $\mathrm{p}(x_0|t) = 1$, when t<0. The initial time-point was set to 1 for all three processes, with $\mathrm{p}(x_0|t) = 1$, when t=1.

Figure 1: The three types of sample space expansion are compared in terms of $p(x_0|t)$

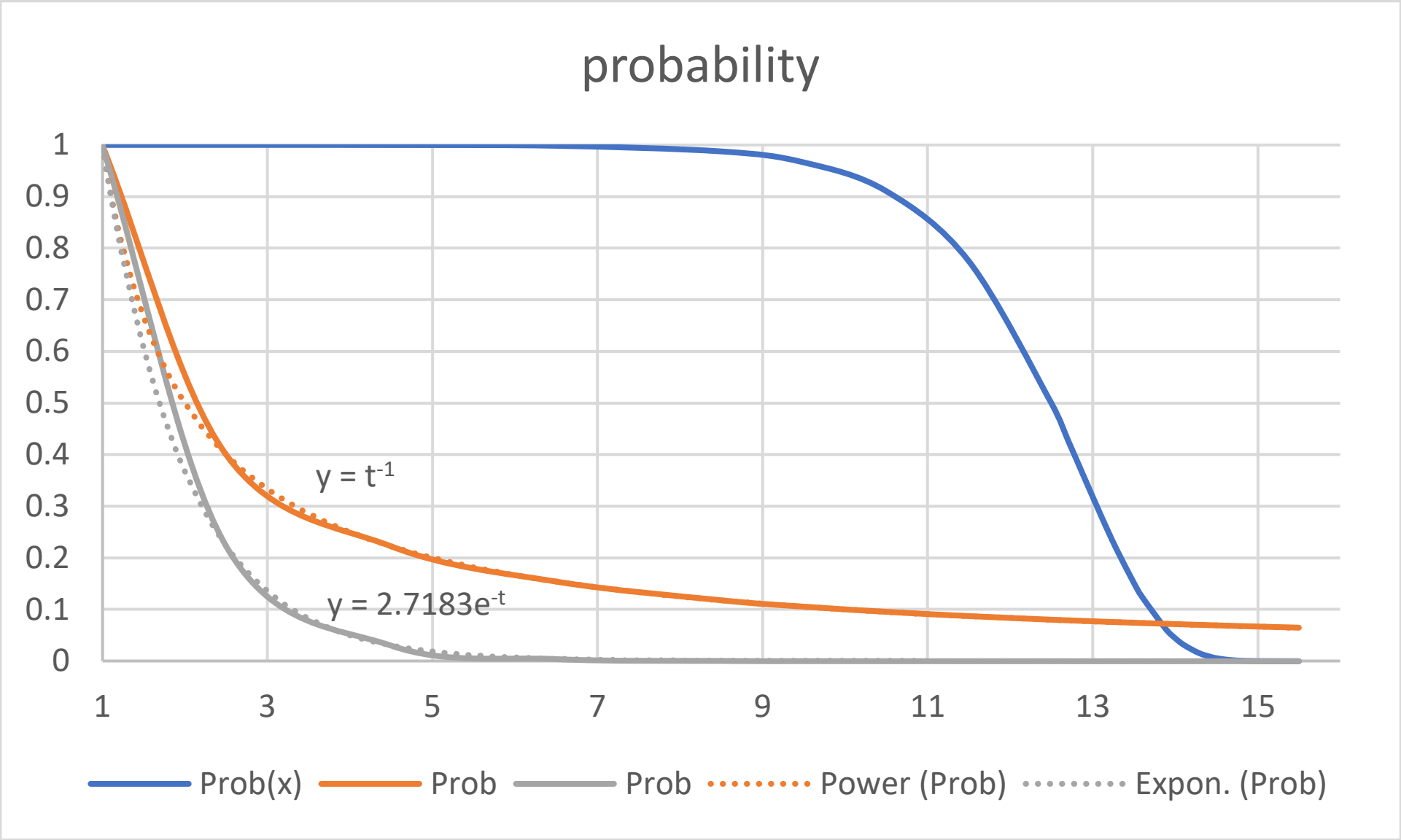

Figure 2: The three types of sample space expansion are compared in terms of $H(t)$

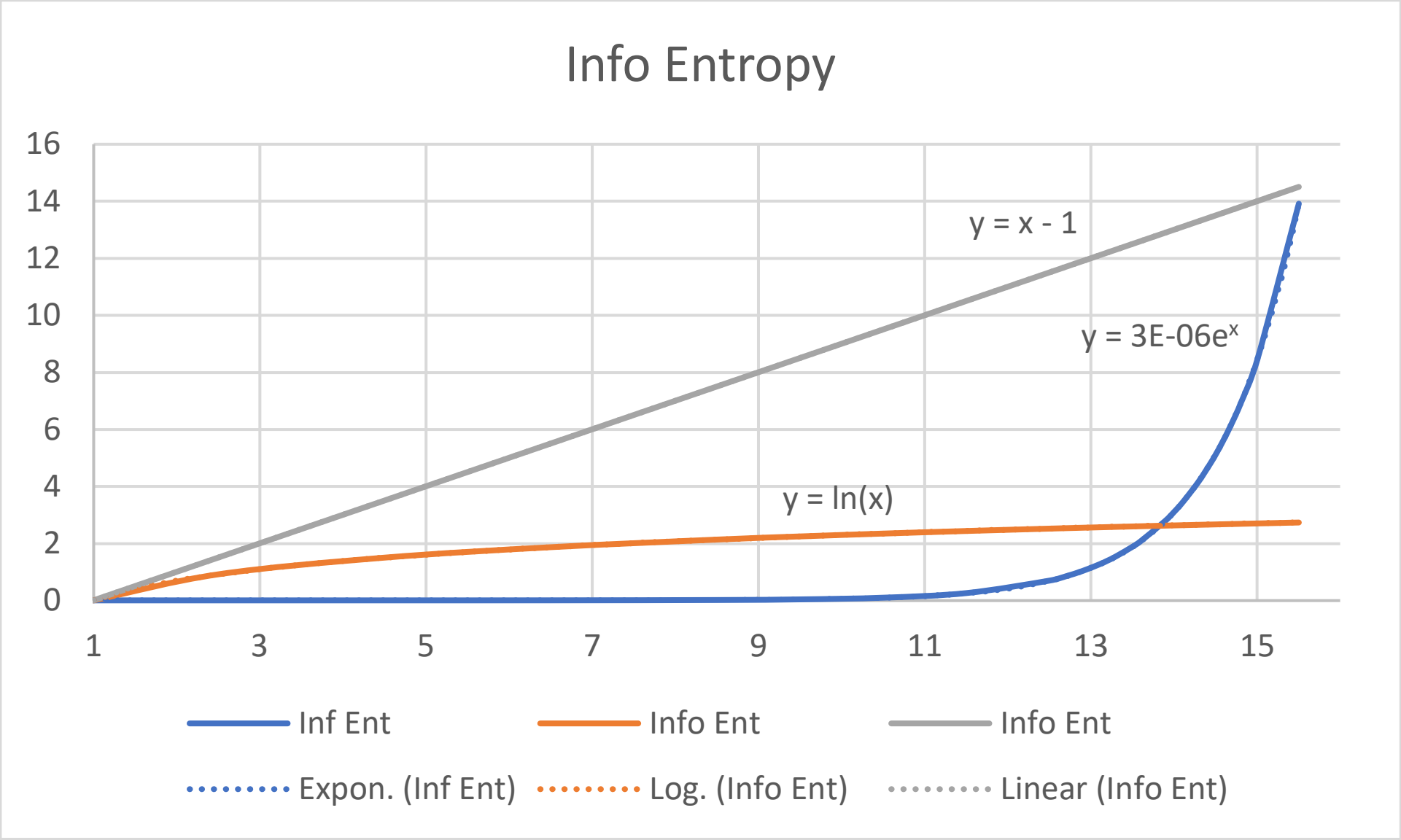

## 3.2 Double exponential expansion of the sample space with time

The double exponential expansion of the sample space with time (from a natural log relationship between t and n, the number of sample space doublings) will now be characterised in more detail, for the case in which $\mathrm{p}(x_0|t) = 1$, when t=1.

The natural log relationship between t and n, the number of sample space doublings is shown in Figure 3.

Figure 3: Natural log relationship between t and n, the number of sample space doublings

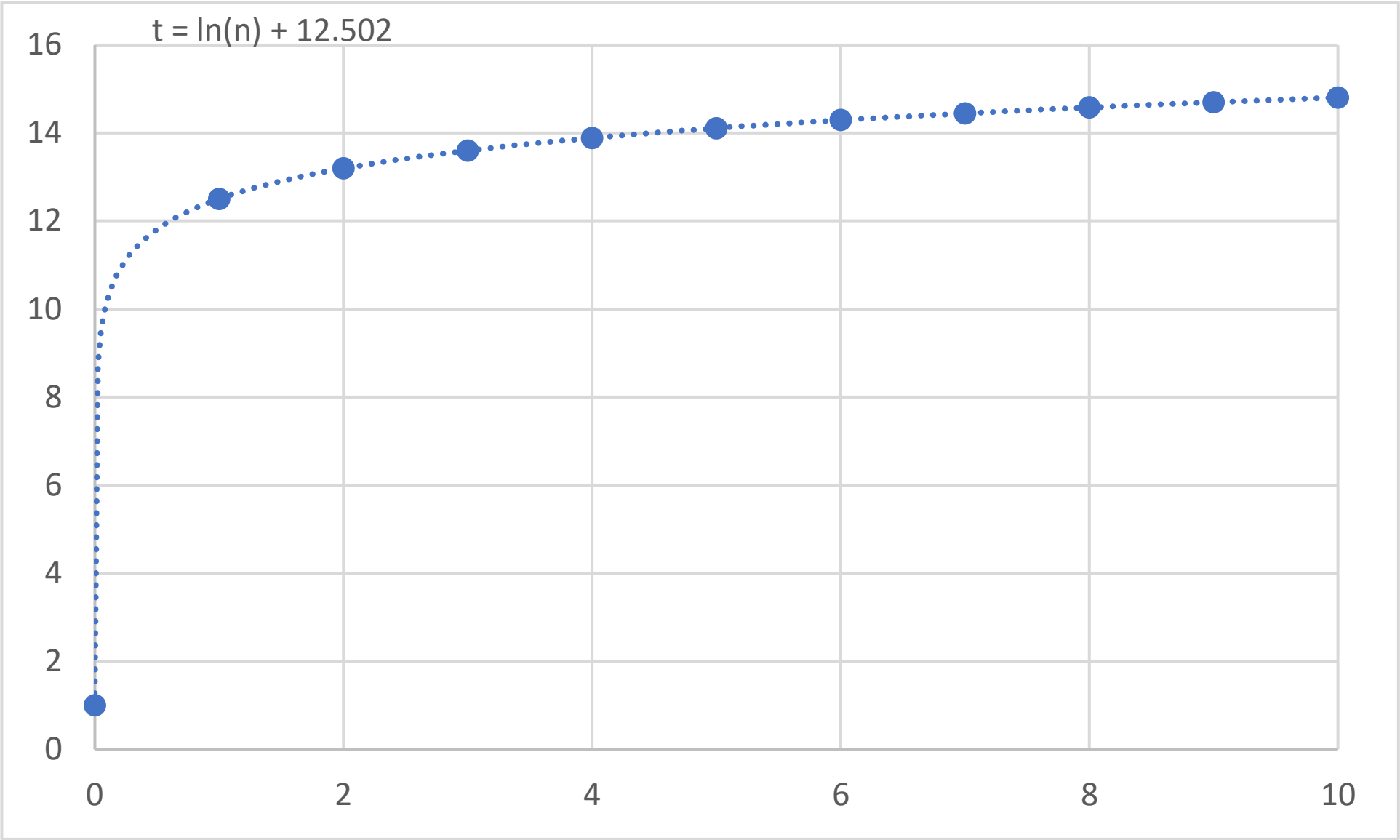

The double exponential expansion of the sample space with time is given in Figure 4. The double exponential nature of the expansion with time can be seen when a linear relationship is obtained with time following a double natural log transformation of the expansion of the sample space with time (Figure 5).

Figure 4: Double exponential expansion of the sample space with time

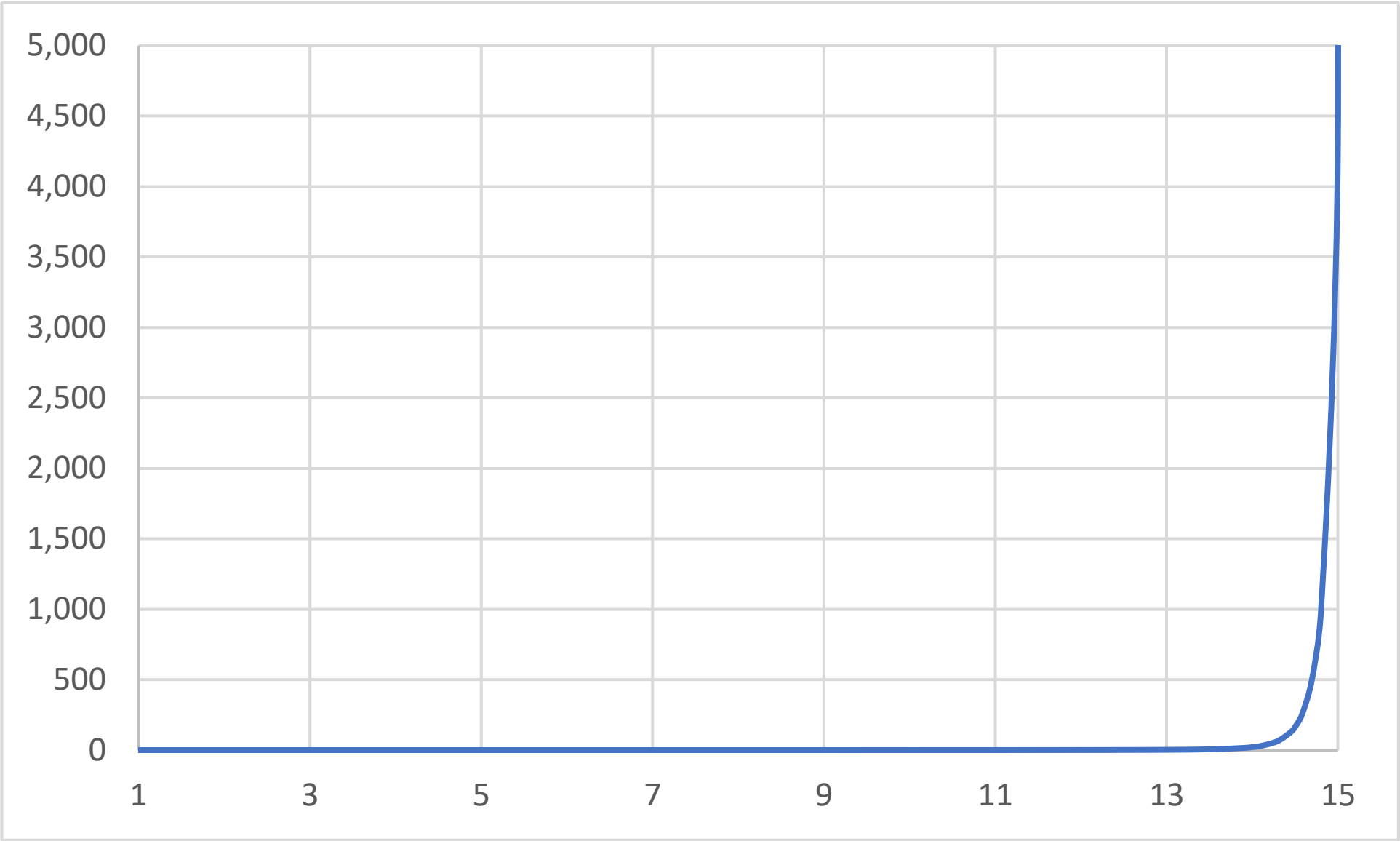

Figure 5: Double natural log transformation of the expansion of the sample space with time

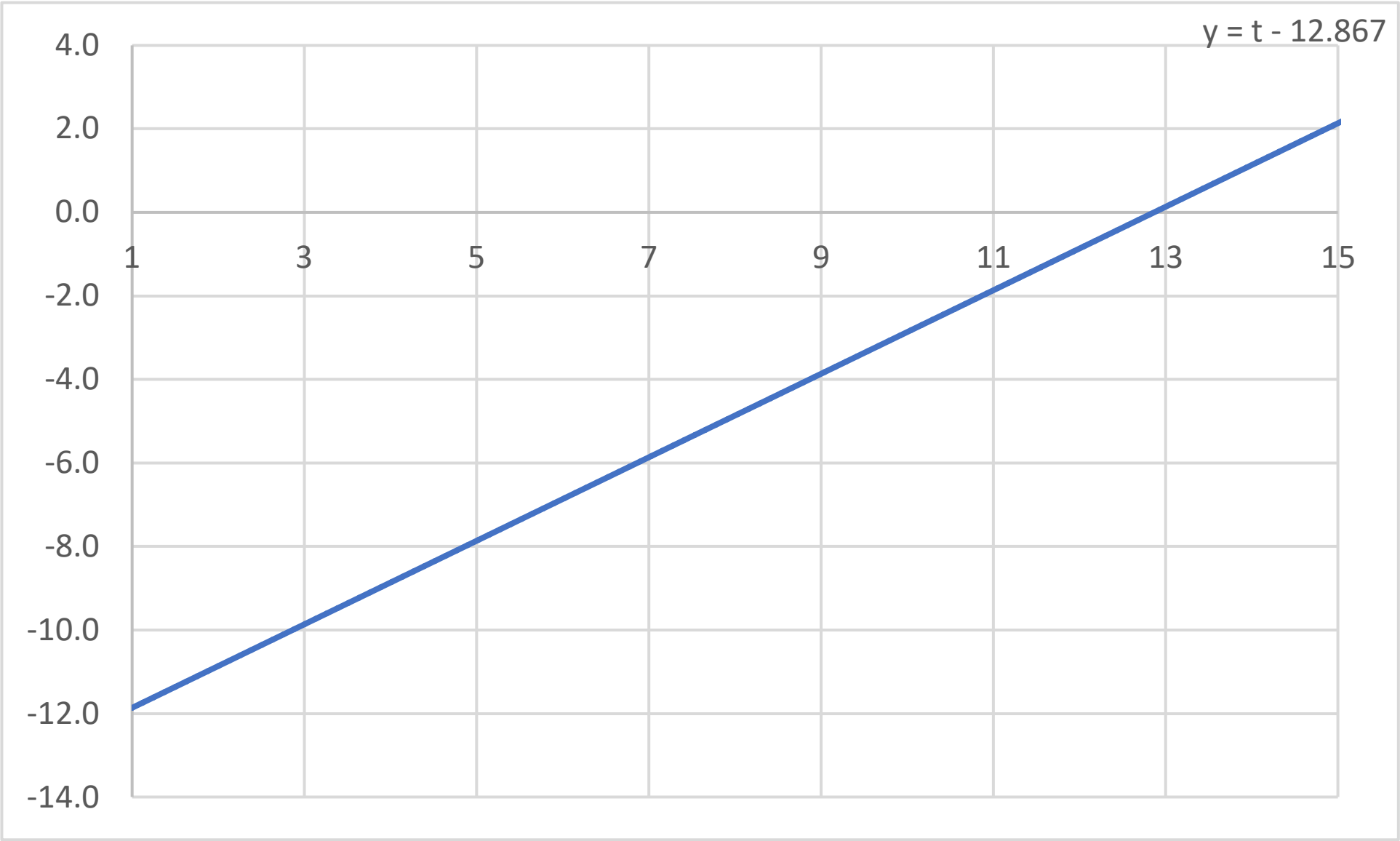

A semi-log plot of $\mathrm{p}(x_0|t)$ and the associated H(t) is given in Figure 6. The exponential growth in the information entropy is demonstrated by the linear increased in the semi-log plot.

Figure 6: Semi-log plot of $p(x_0|t)$ and the associated H(t)

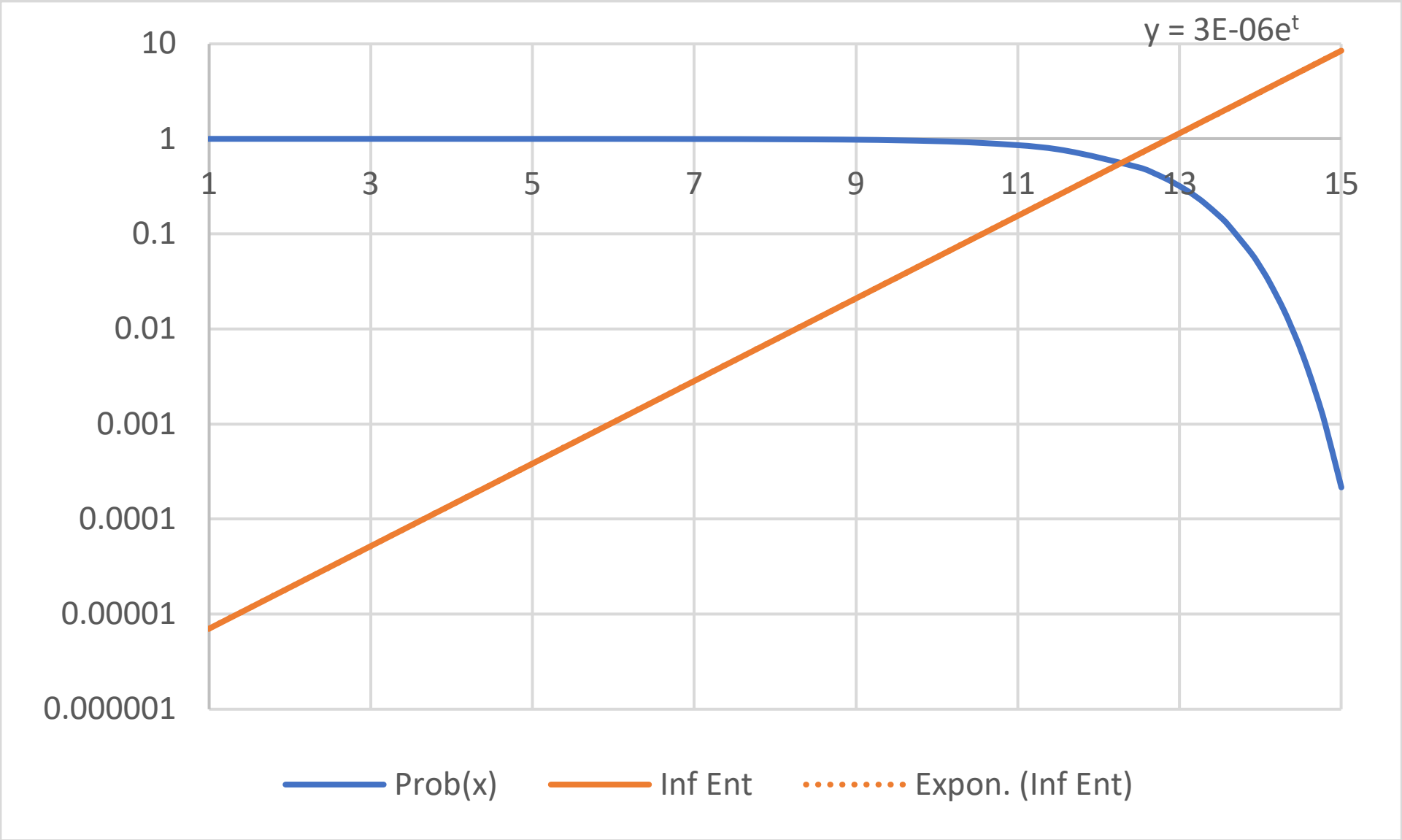

## 4. Discussion

The probability and information entropy of a process with an increasing sample space by means of different functional forms of expansion has been characterised in the case of (1) a mono-exponential expansion of the sample space; (2) a power function expansion; (3) double exponential expansion. What determines the different functional forms of the expansion of the sample space is the functional relationship between time and n (no. of sample space doublings). This is summarised in Table 1.

If the expansion of the sample space occurred simultaneously via the three independent processes (P1, P2, P3), the combined overall process could be determined by the following two equations:

$$\mathrm{p}(x_0|t) = \mathrm{p}(x_0|P1, t)\ x\ \mathrm{p}(x_0|P2, t)\ x\ \mathrm{p}(x_0|P3, t)$$

and

$$\mathrm{H(t) = H(t|P1) + H(t|P2) + H(t|P3)}$$

The double exponential expansion of the sample space with time (from a natural log relationship between t and n) is of particular interest because it describes a "hyperinflationary" process. In terms of the fiat money supply, such an expansion would lead to hyperinflation, such that the "money" would become increasingly "worthless" with time ($\mathrm{p}(x_0|t)$ approx. $= 0$, Figures 1 and 6). Such a model might be of value to better understand the factors resulting in fiat currency hyperinflation [2].

Historical hyperinflation of fiat currencies has been studied [3]. In particular, the double exponential nature of extreme cases of hyperinflation has also been investigated [4]. The hyperinflation of in the Weimar Republic between the period November 1922 to October 2023 is investigated in this paper. This case of hyperinflation was found to be double exponential in nature [4]. The data analysed in this paper are based on the value of one gold Mark in paper Marks over the period from the middle of 1922 to the end of 1923 (obtained from [5]) and are plotted in Figure 7.

Figure 7: Semi-log plot of the value of one gold Mark in paper Marks over the period from the middle of 1920 to the end of 1923 (in months), with time t=0 being the middle of 1920 (obtained from [5]).

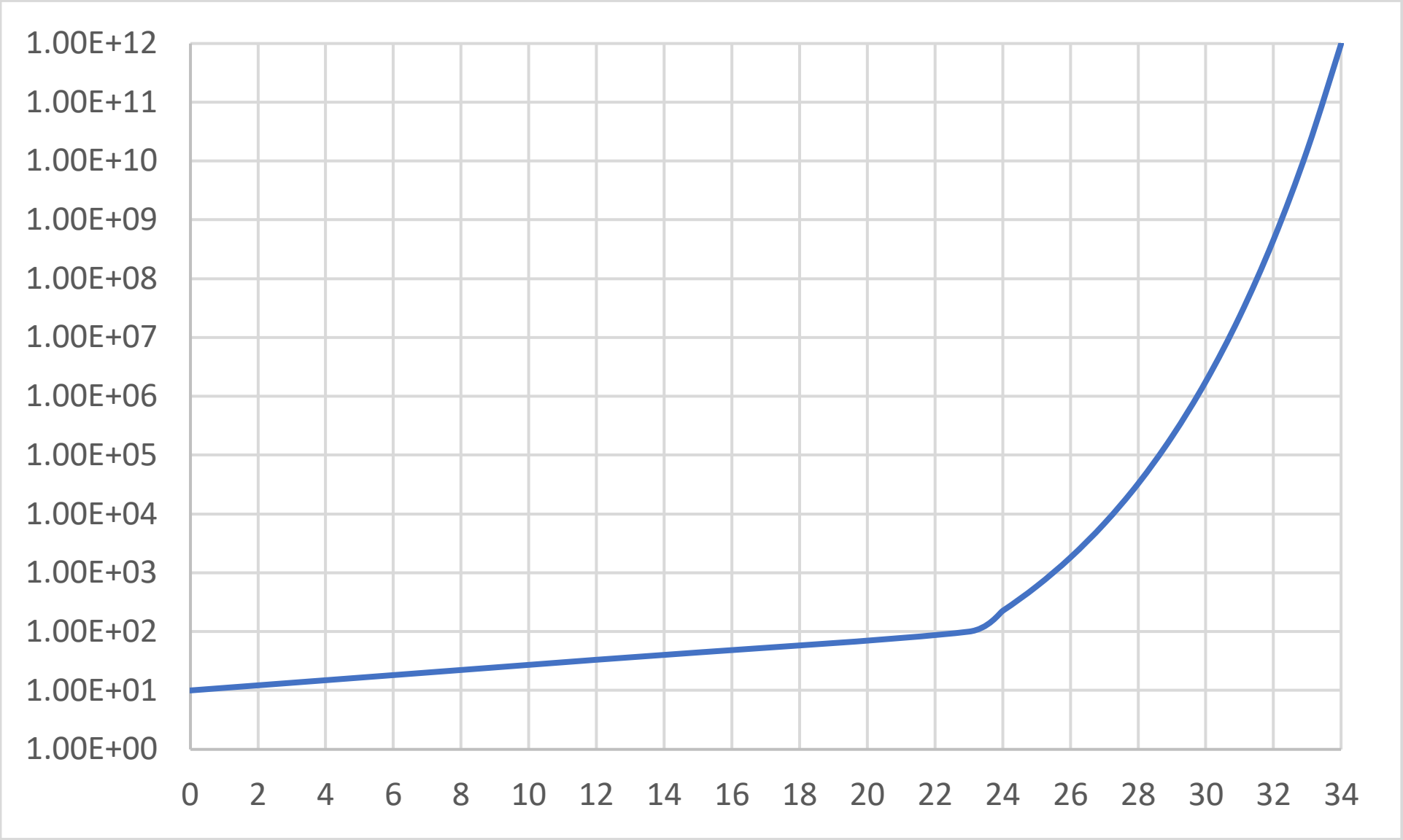

For the 24 months up to November 1922 (t=23), the number of paper Marks required to purchase one gold Mark, $m_1(t)$, is given approximately by:

$$m_1(t) = 10\ x\ \exp\ (0.1001\ x\ t)$$

For the following 12 months until the end of October 1923, $m_2(t)$, is given approximately by:

$$m_2(t) = \ 10\ x\ \exp\ (0.1001\ x\ 23)^{\exp\ (\lambda\ x\ (t-23))}$$

where, λ= 0.1629

The double exponential nature of the hyperinflation from the period November 1922 until the end of October 1923 is shown by Figure 8.

Figure 8: Plot of the double natural log of the number of paper Marks required to purchase one gold Mark, from the period November 1922 until the end of October 1923 (with t=23 being November 1922)

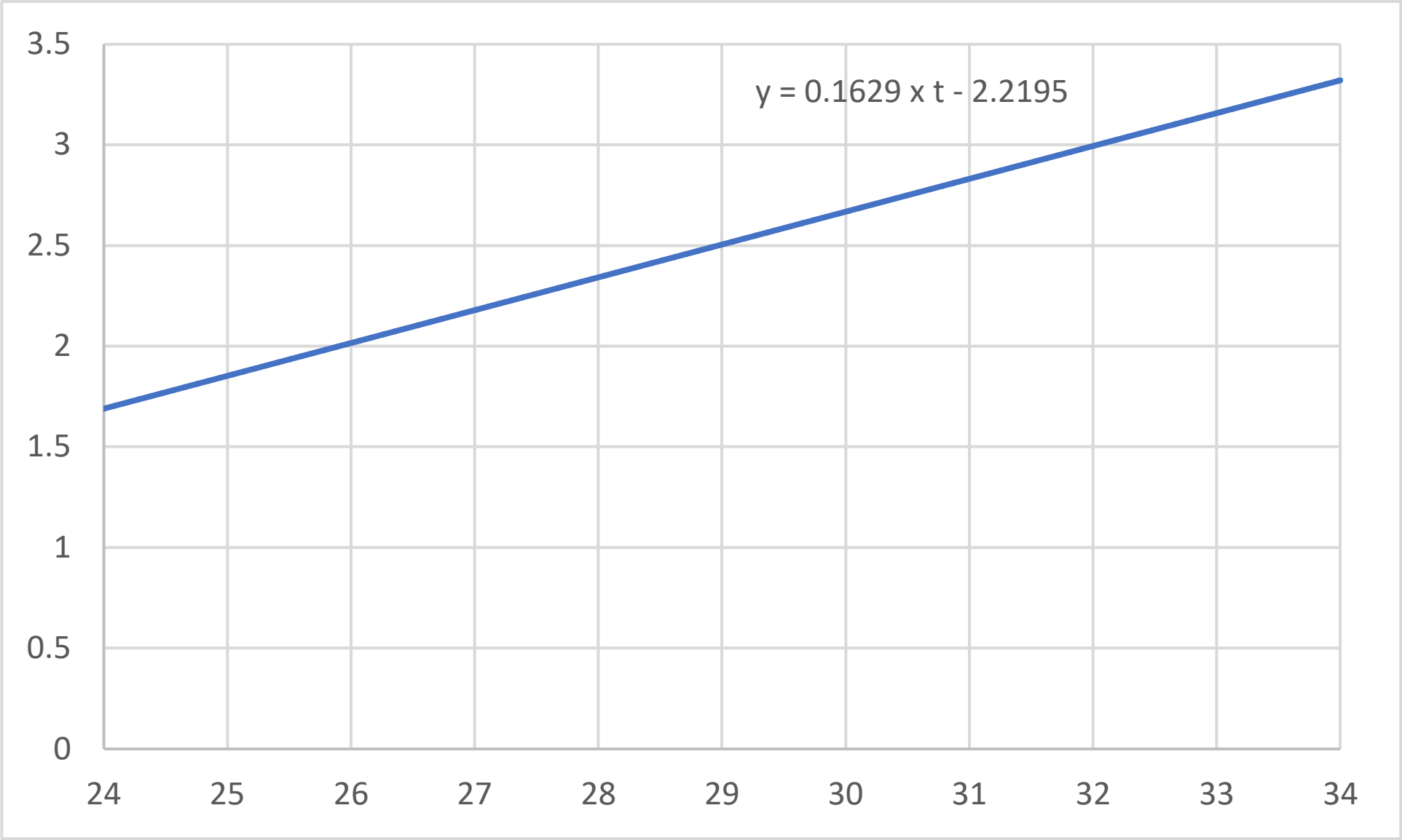

Over this time period, the purchasing power of the paper Mark to purchase one gold Mark is the inverse of the equations given above. Towards the end of 1923, the purchasing power of a unit of the paper Mark becomes close to zero, i.e., the unit of currency became "worthless" (1 paper Mark = $10^{-12}$ gold Mark). This is analogous to the declining probabilities with time given in Figure 1. From the purchasing power of the paper Mark to purchase one gold Mark, the information entropy of this hyperinflationary process can be determined and is given in Figure 9.

Figure 9: The information entropy of the purchasing power of the paper Mark over the period from the middle of 1920 to the end of 1923 (in months, where t=23 is November 1922).

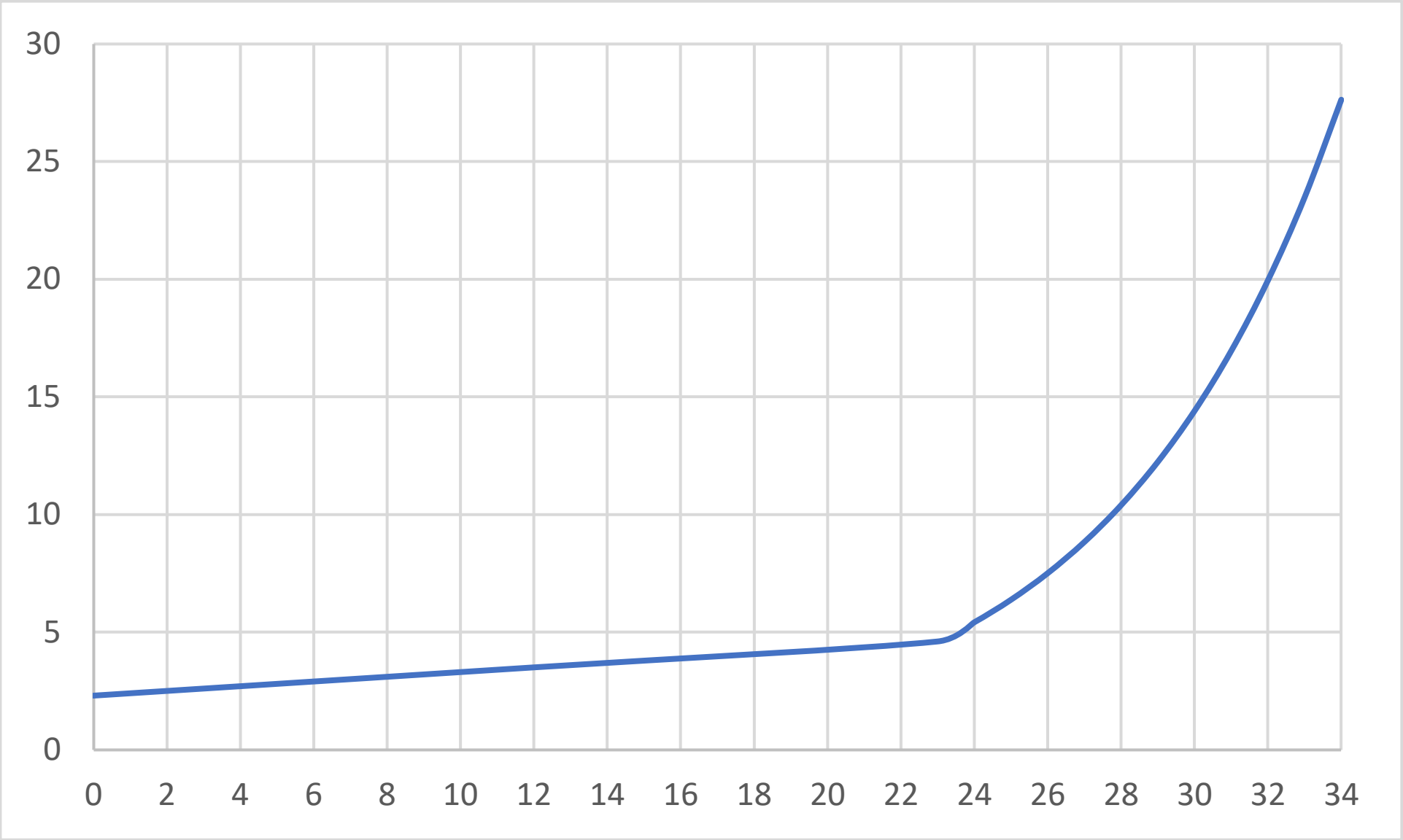

## 5. Conclusion

The probability and information entropy of a process with an increasing sample space by means of different functional forms of expansion has been characterised. If the expansion of the sample space occurred simultaneously via three (or more) independent processes, of different functional forms, the combined overall process can also be determined. The double exponential expansion of the sample space with time (from a natural log relationship between time and n (no. of sample space doublings) is of particular interest because it describes a "hyperinflationary" process. Over the period from the middle of 1920 to the end of 1923, the purchasing power of the Weimar Republic paper Mark to purchase one gold Mark became close to zero (1 paper Mark = $10^{-12}$ gold Mark). From the purchasing power of the paper Mark to purchase one gold Mark, the information entropy of this hyperinflationary process was determined.

## Supplementary materials

There are no supplementary materials.

## Acknowledgements

The author gratefully acknowledges those academic and non-academic people who kindly provided feedback on earlier drafts of this paper. No financial support was received for any aspect of this research.